\documentclass[aoas,MSNbibl,nameyear,dvips]{arximspdf}
\usepackage{graphicx}

\doi{10.1214/10-AOAS428}
\volume{5}
\issue{2A}
\pubyear{2011}
\firstpage{628}
\lastpage{644}

\makeatletter
\renewcommand{\epsilon}{\varepsilon}
\renewcommand{\cite}{\citet}

\newtheorem{theorem}{Theorem}

\newcommand{\ex}{\mathbf{x}}
\makeatother

\begin{document}
\begin{frontmatter}

\title{Structured, sparse regression with application to HIV drug resistance}
\runtitle{Structured sparsity}

\begin{aug}
\author[A]{\fnms{Daniel} \snm{Percival}\corref{}\thanksref{t1}\ead[label=e1]{dperciva@stat.cmu.edu}},
\author[A]{\fnms{Kathryn} \snm{Roeder}\ead[label=e2]{roeder@stat.cmu.edu}},
\author[B]{\fnms{Roni} \snm{Rosenfeld}\ead[label=e3]{Roni.Rosenfeld@cs.cmu.edu}}\break
\and
\author[A]{\fnms{Larry} \snm{Wasserman}\ead[label=e4]{larry@stat.cmu.edu}}
\thankstext{t1}{Supported in part by the National Institutes of Health SBIR Grant 7R44GM074313-04 at~Insilicos~LLC.}
\runauthor{Percival, Roeder, Rosenfeld and Wasserman}
\affiliation{Carnegie Mellon University}
\address[A]{D. Percival\\
K. Roeder\\
L. Wasserman\\
Department of Statistics\\
Carnegie Mellon University\\
Pittsburgh, Pennsylvania 15213\\
USA\\
\printead{e1}\\
\phantom{E-mail:\ }\printead*{e2}\\
\phantom{E-mail:\ }\printead*{e4}} %adresu isvedimo komanda gale!
\address[B]{R. Rosenfeld\\
School of Computer Science\\
Carnegie Mellon University\\
Pittsburgh, Pennsylvania 15213\\
USA\\
\printead{e3}}
\end{aug}

% HISTORY:
\received{\smonth{2} \syear{2010}}
\revised{\smonth{9} \syear{2010}}

% ABSTRACT
\begin{abstract}
We introduce a new version of forward stepwise regression.  Our
modification finds solutions to regression problems where the selected
predictors appear in a structured pattern, with respect to a
predefined distance measure over the candidate predictors.  Our method
is motivated by the problem of predicting HIV-1 drug resistance from
protein sequences.  We find that our method improves the
interpretability of drug resistance while producing comparable
predictive accuracy to standard methods.  We also demonstrate our
method in a simulation study and present some theoretical
results and connections.
\end{abstract}

% KEYWORDS
\end{frontmatter}

%s1 ###
\section{Introduction}

About twenty antiretroviral drugs are currently available for the
treatment  of human immunodeficiency virus type 1 (HIV-1).  The great
majority of these function by inhibiting the activity of various
proteins produced by the HIV-1 virus, effectively impairing the virus'
ability to reproduce.  Resistance to these drugs develops when a
mutation changes the structure of the target protein enough to
frustrate the drug while still maintaining the function of the protein.
HIV-1 is capable of rapid mutation, and is thus often able to adapt to
antiretroviral therapy.  Understanding the genetic basis for this
developed resistance would allow more effective development of new
drugs, as well as more informed prescription of the currently available
drugs.

Sequencing HIV-1 proteins can be done reliably, and well-designed
in-vitro experiments are available for testing the resistance of a
particular strain of HIV-1 to drugs; see \cite{assay1} and
\cite{assay2}. We approach this problem using regression. This problem
setting leads us to build models to predict drug resistance using
mutations in the amino acid sequence of the target proteins.  We desire
models that are easy to interpret and take into account properties of
proteins and amino acids.  In particular, it is well known that
proteins generally function using areas called active sites, that are,
simply, areas of the sequence where the protein binds or otherwise
interacts with other molecules.  This fact leads us to believe that
important mutations will tend to be clustered around such sites.

Protein sequences can be thought to have two layers of structure: the
primary sequence consisting of a single string of adjacent amino acids,
and a~secondary structure created by protein folding.  We can measure
the distance between amino acids in a protein sequence roughly using
the differences in position in the primary sequence.  When the
protein's folding structure is known, three-dimensional distance can be
calculated for any two amino acid positions.  But even when the
structure of the protein is unknown, because of the continuity of the
primary sequence, clustering in three-dimensional space generally
corresponds to clustering in the protein primary sequence.\looseness=1

We therefore build models for predicting resistance from mutations that
have the following two properties: (1) \textit{Sparsity}---a model that
uses only a few mutations is easier to interpret and apply.  (2)
\textit{Structure}---following the concept of active sites, we wish to
use mutations that are clustered in the protein primary sequence.  Note
that this second property is desirable in other applications.  For
example, \cite{pathways} use genetic pathways to model the genetic
influences on prostate cancer. These pathways can be modeled as a
structure on individual genes.  In this paper we introduce a variable
selection method that builds regression models that satisfy these two
properties.

Forward stepwise regression and the lasso are two popular automatic
variable selection techniques that are effective at finding sparse
regression models.  Given data $(X_1,Y_1),\ldots, (X_n,Y_n)$ where
$Y_i\in\mathbb{R}$ and $X_i\in\mathbb{R}^p$, the lasso~%
$\hat{\beta}_{\mathrm{lasso}}$ estimator due to
\cite{Tibshirani94regressionshrinkage} minimizes
%
%e1 ###
\begin{equation}
\sum_{i=1}^n (Y_i - X_i^T\beta)^2 + \lambda \Vert\beta\Vert_1,
\end{equation}
where $\Vert\beta\Vert_1 = \sum_j | \beta_j|$ and $\lambda>0$ is a tuning
parameter which controls the amount of regularization. Forward stepwise
regression is a greedy method that adds one predictor, that is, one
element $X_i$, at a time. Both produce sparse solutions, meaning that
$\hat\beta_j=0$ for most $j$. Sparse solutions are attractive both
computationally and for interpretation.

Recent results show that both methods yield estimators with good
properties. See  \cite{bunea}, \cite{greenshtein}, \cite{wain} for
results on the lasso, and \cite{barron} for results on forward stepwise
regression. These papers show that, under weak conditions, both
approaches yield predictors that are  $O(n^{-1/4})$ close to the
optimal sparse linear predictor. Moreover, this rate cannot be
improved.  In our application, extra information is available---we
expect nonzero $\beta_j$'s to cluster together. In this case, we would
like to add an additional constraint to the regression.

In this paper we introduce a modification of forward stepwise
regression that encourages the selection of new predictors that are
``close''---with respect to a distance measure over the predictors---to
 those already included in the model.  We show that our method,
Clustered and Sparse Regression (CaSpaR), is useful in regression
problems where we desire both a sparse and structured solution.\vspace*{3pt}

%s2 ###
\section{Data}

The Stanford HIV drug resistance database described in Rhee et al.~(\citeyear{database})
is a large data set of HIV-1 protease sequences, along with resistance
phenotypes for up to seven different protease inhibitor (PI) drugs for
each sequence. This database is a combination of smaller data sets
collected in different clinical trials.  Since both the genotyping and
phenotyping experiments are well standardized, such a joining of data
will not give rise to significant heterogeneity-in-sample concerns.
Each protease protein sequence is 99 amino acids long.  The phenotypes
are obtained from in-vitro experiments, and are measured in terms of
number of multiples of standard dose of drug needed to suppress virus
reproduction.

We can cast the problem of connecting genotype to phenotype as a
regression problem by treating each mutation as a predictor.  Previous
studies by  Rhee et al.~(\citeyear{Soo}) and  \cite{g2p} have used most modern sparse
regression and classification techniques to attack this problem.  We
seek a model that will take into account protein active sites.\vspace*{3pt}

%s3 ###
\section{CaSpaR}

We first introduce the usual regression setting.  We have an $n \times
p$ data matrix $\mathbf{X}$ and $n\times 1$ response vector
$\mathbf{Y}$. We use the usual linear model\vspace*{2pt}
%
%e2 ###
\begin{equation}
\mathbf{Y} = \mathbf{X}\beta + \epsilon.\vspace*{2pt}
\end{equation}
Define the support of $\beta$ by\vspace*{2pt}
%e3 ###
\begin{equation}
\operatorname{supp}(\beta) = \{ j\dvtx \beta_j \neq 0, j = 1,\ldots,p \}.\vspace*{2pt}
\end{equation}
We assume that $\beta$ is sparse (most $\beta_j$'s are 0) and also that
$\operatorname{supp}(\beta)$ has structure. We base this structure on a
distance measure $d(\cdot,\cdot)$ over the set of predictors:\vspace*{2pt}
%
%e4 ###
\begin{equation}
d(\cdot,\cdot)\dvtx \{1,\ldots,p\} \times \{1,\ldots,p\} \to \mathbb{R}.\vspace*{2pt}
\end{equation}
Specifically, we assume that the nonzero elements of $\beta$ are
spatially clustered with respect to $d(\cdot,\cdot)$.  In other words,
the nonzero entries of $\beta$ appear in some number of groups in which
the members are ``close'' to each other---as defined by
$d(\cdot,\cdot)$.  Our goal is to accurately recover $\beta$, with
particular emphasis on this sparsity structure.

%t1 ###
\begin{table}
\caption{Forward stepwise regression} \label{FSR}%
%\vspace*{-3pt}
\begin{tabular}{@{}cl@{}}
\hline
1. & Input: $A = \varnothing$, $\mathbf{X}$, $\mathbf{Y}$, $\epsilon >
0$.\\
2. & Fit an OLS model: $\hat{\beta} =
  \arg\min_\beta \Vert\mathbf{X}\beta - \mathbf{y}\Vert^2_2$, s.t.
  $\operatorname{supp}(\beta) \subseteq A$.\\
3. & Set $i^* = \arg \max_{\{i \notin A\}} |(\mathbf{X}\beta -
  \mathbf{y})^T\mathbf{x}_i|$.\\
4. & If $|x^T_{i^*}(\mathbf{X}\beta - \mathbf{y})| < \epsilon$ then
  stop, else set $A = A \cup i^*$ and go to step 2.\\
  \hline
\end{tabular}
%\vspace*{-12pt}
\end{table}

%t2 ###
\begin{table}[b]

\caption{CaSpaR: Clustered and Sparse Regression} \label{CaSpaR}%
%\vspace*{-3pt}
\begin{tabular}{@{}cl@{}}
\hline
1. & Input: $A = \varnothing$, $\mathbf{X}$, $\mathbf{Y}$, $h>0$,
  $\alpha \in (0,1)$, $\epsilon > 0$.\\
2. & Fit an OLS model: $\hat{\beta} =
  \arg\min_\beta \Vert\mathbf{X}\beta - \mathbf{y}\Vert^2_2$, s.t.
  $\operatorname{supp}(\beta) \subset A$.\\
3. & $\forall l \notin A$, calculate: $W_l = \frac{1}{|A|} \sum_{\{k
\in A\}} K_h(d(l,k))$.  If this is the first iteration
  of the\\
  & algorithm, set $W_l = 1$, $\forall l$.\\
4. & Set $l^* = \arg \max_{\{l \notin A\}} W_l|(\mathbf{X}\beta -
  \mathbf{y})^T\mathbf{x}_l|$.\\
5. & If $|x^T_{l^*}(\mathbf{X}\beta - \mathbf{y})| < \epsilon$ then
  stop, else set $A = A \cup l^*$ and go to step 2.\\
  \hline
\end{tabular}
\end{table}

We want to modify a sparse regression technique to produce solutions
with clusters of nonzero coefficients.   Penalized techniques such as
the lasso are difficult to modify for this purpose. Recall that the
lasso finds $\hat\beta$ that minimizes
%
%e5 ###
\begin{equation}
Q(\beta)=\sum_{i=1}^n (Y_i -  \mathbf{X}_i^T\beta)^2 + \lambda \sum_j
|\beta_j|.
\end{equation}
The lasso is computationally efficient because $Q(\beta)$ is convex. It
is difficult to add a penalty to $Q(\beta)$ that encourages clustered
solutions while maintaining convexity. Note that the \textit{fused lasso} due
to \cite{tibshirani} adds a~penalty of the form
$\sum_j |\beta_j-\beta_{j-1}|$. This forces nearby coefficients to be
close together in sign and magnitude. We want the support points to be
close together, but we do not want to force the values of the
coefficients to be close together. Instead, we are only concerned with
the inclusion or exclusion of predictors.

Stepwise procedures are more flexible and easier to modify, because we
do not need to worry about maintaining the convexity of an objective
function.  We therefore propose a modification to forward stepwise
regression (see Table \ref{FSR} for a~description of forward stepwise
regression).  We call our algorithm \textit{CaSpaR} (Clustered and Sparse
Regression); see Table \ref{CaSpaR}.\vfill\eject

In each iteration of forward stepwise regression, the following
quantities are used to select the next predictor to be added to the
model:
%
%e6 ###
\begin{equation}
C_j = |(\mathbf{X}\beta - \mathbf{Y})^T\mathbf{x}_j|,
\end{equation}
where $\mathbf{x}_j$ denotes the $j$th column of $\mathbf{X}$.  Note
that the $C_j$ are proportional to the correlations between each
candidate predictor and the current residuals if the columns of
$\mathbf{X}$ are scaled to empirical mean zero, variance one.  We wish
to encourage the selection of predictors that are close, with respect
to $d(\cdot,\cdot)$, to those already in the model.  To do this, we
multiply the $C_j$ by a~kernel, which we construct based on the current
active set $A$.  This kernel will weight the $C_j$ so that predictors
that are close to those already in the model receive more weight than
those that are not.

%f1 ###
\begin{figure}[b]

\includegraphics{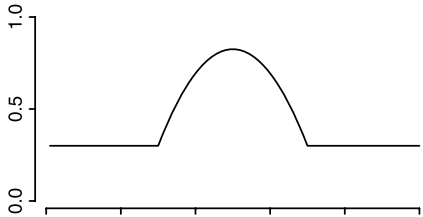}

\caption{The Stetson kernel, with an Epanechnikov kernel.}\label{stet}
\end{figure}

Formally, suppose we have a kernel $K_h$ that is centered at 0, where
$h$~de\-notes the bandwidth parameter.  Then, for all $l \notin A$, we
compute
%
%e7 ###
\begin{equation}
W_l = \frac{1}{|A|} \sum_{\{l \in A\}} K_h(d(l,k)).
\end{equation}
We then select the next predictor $j^*$ using a weighted criterion:
$W_j(\mathbf{X}\beta - \mathbf{y})^T\mathbf{x}_j$.  For most familiar
kernels, such as a Gaussian kernel or an Epanechnikov kernel, this has
the effect of boosting the criterion for predictors ``near'' those
already included in the model, and diminishing the criterion for those
``far away.'' For practical application, we recommend a mixture of a
familiar kernel, such as a boxcar or Epanechnikov, and a uniform
distribution.  This mixture, which we call the Stetson kernel,
introduces an additional mixing parameter $\alpha$:
%
%e8 ###
\begin{equation}
K_{h,\alpha}(x) = \alpha + (1-\alpha)K_h(d(x)),
\end{equation}
where $K_h$ is a kernel such as a boxcar, Epanechnikov or Gaussian. An
example of this kernel appears in Figure \ref{stet}.  We particularly
recommend the Epanechnikov or the boxcar kernel, because these kernels
have no impact at all on predictors outside their bandwidth, and so
$W_i = \alpha$ for these predictors.  While this usually makes no
difference in predictor selection, it simplifies precise computation
and interpretation.

The advantage of the Stetson kernel is that this mixture allows
multiple groups of predictors to appear in the sparsity structure.  If
we were instead to only use a familiar kernel, then we  would have $W_j
= 0$ (or extremely small) for those $j$ far enough away from predictors
already included in the model.  This approach would lead to only a
single group in the sparsity structure, built around the first selected
predictor, whereas most applications call for multiple groups.  The
Stetson kernel avoids this problem.  The uniform part of the Stetson
kernel allows new predictor groups to enter the model.  The kernel part
of the mixture encourages clustering around predictors already included
in the model.

Finally, note that CaSpaR is closely related to forward stepwise
regression. Indeed, with $\alpha=1$ CaSpaR reduces to forward stepwise
regression.  Therefore, as long we consider $\alpha = 1$ when picking
parameters, we always consider the forward stepwise regression
solution.  Consequently, we have a~loose guarantee that CaSpaR does no
worse than forward stepwise regression. Moreover, we expect that some
theoretical results relating to forward stepwise regression can be
adapted to CaSpaR.

%s3.1 ###
\subsection{Tuning parameters}

$\!\!$CaSpaR has three tuning parameters: $\epsilon, h$ and~$\alpha$. The
parameter $\epsilon$ controls the sparsity of the fitted model. The
parameters $h$ and $\alpha$ control the amount of structure in the
estimated support. For the Stetson kernel, as the bandwidth $h$
decreases, the predictors become more tightly grouped.  As $\alpha$
increases, new clusters are allowed to form more easily.  In the
special case where $\alpha = 1$, the method reduces to the usual
forward stepwise regression method.  Let $\mathit{CV}(\epsilon,h,\alpha)$ denote
the cross-validation score. We choose the parameters by minimizing
$\mathit{CV}(\epsilon,h,\alpha)$.  Note that since small changes in $h$ or
$\alpha$ do not affect the order of predictor selection, this tuning
can be accomplished using a simple grid search.

%s4 ###
\section{Results}

We now return to our application to HIV drug resistance.  Our data set
consists of 553 amino acid sequences, all 99 amino acids in length.
Each amino acid sequence corresponds to a different strain of HIV found
within a patient.  Each sequence has resistance measurements for up to
seven HIV inhibiting drugs.  Thus, the number of sequences available
for our analysis varies depending on which drug we consider.

After we choose a drug and take the appropriate subset of our 553
sequences, we create our predictors.  With twenty known amino acids,
each position in these sequences thus takes twenty possible values.  We
thus define our mutation predictors as follows. At each of the 99
positions, we first search across all of the available sequences and
record the set of amino acids that appear at that position in the data.
This set is the collection of possible mutations at that particular
position.  If there is only one amino acid in this set, this
corresponds to the case where that particular position displays no
variation in amino acid over the data, and thus can be dropped from the
analysis.  We use mutations from positions with $M>1$ possible amino
acids to create $M-1$ predictors.  Each of these predictors is an
indicator variable which, for a particular sequence, is equal to $1$ if
the particular amino acid appears at that particular position and $0$
otherwise.  We refer to these predictors as mutations.  Since each
mutation has an associated position in the primary sequence, we can
define a distance between predictors as the absolute difference of
their positions.  Thus, the mutations that occur at the same position
are distance 0 from each other.

%t3 ###
\begin{table}[b]
  \caption{Summary of results across all models and drugs.  For each
    model, we give the mean-square-error, as well as the number of
    mutations (predictors) selected in parentheses.  We see that
    CaSpaR is comparable to forward stepwise regression in terms of
    MSE, with about the same number of predictors included in the
    model.  The lasso does better in MSE, but includes many more
    mutations than either stepwise method.  As we previously noted,
    neither forward stepwise regression nor the lasso allows for a
    structured sparse solution}\label{HIV}%
    \begin{tabular*}{\textwidth}{@{\extracolsep{\fill}}lccc@{}}
\hline
\textbf{Drug name} & \textbf{Stepwise} & \textbf{CaSpaR} & \textbf{Lasso} \\
\hline
 APV & 0.514  (7)\phantom{0}& 0.477  (14)& 0.422  (51) \\
 ATV & 0.588  (6)\phantom{0}& 0.494  (11)& 0.477  (39) \\
 IDV & 0.541  (13)& 0.580  (10)& 0.449  (77) \\
 LPV & 0.614  (5)\phantom{0}& 0.507  (15)& 0.518  (35) \\
 NFV & 0.650  (19)& 0.637  (22)& 0.661  (40) \\
 RTV & 0.659  (8)\phantom{0}& 0.714  (5)\phantom{0}& 0.570  (58) \\
 SQV & 0.426  (31)& 0.508  (21)& 0.447  (63) \\
\hline
\end{tabular*}
\end{table}

Our design matrix $\mathbf{X}$ is thus an $n_{\mathrm{drug}} \times
p_{\mathrm{drug}}$ matrix.  Here $n_{\mathrm{drug}}$ is the number of
sequences with measurements of the resistance score for the drug of
interest.  The number of sequences with resistance measurements for
each drug are as follows:  453 for drug APV,  212 for ATV, 496 for IDV,
300 for LPV, 510 for NFV, 465 for RTV, and 493 for SQV.  We then create
the $p_{\mathrm{drug}}$ mutation indicator predictors as described above.
Since the number of samples varies with the drug, so does the number of
mutation predictors.  The number of predictors for each drug are as
follows: 210 for drug APV, 180 for ATV, 215 for IDV, 199 for LPV, 219
for NFV, 215 for RTV, and 218 for SQV.\looseness=-1

We compare CaSpaR to forward stepwise regression and lasso models. For
all methods, we use ten-fold cross-validation to choose all the tuning
parameters. We use the R package {\textsf{glmnet}}
[\cite{Friedman2010}] to obtain the lasso solution. For CaSpaR, we use
the Stetson kernel and perform a grid search over $\alpha =
\{0,0.1,0.2,\ldots,1\}$, and over $h = \{1,2,3,4\}$ to find the optimal
tuning parameters.

%f2 ###
\begin{figure}

\includegraphics{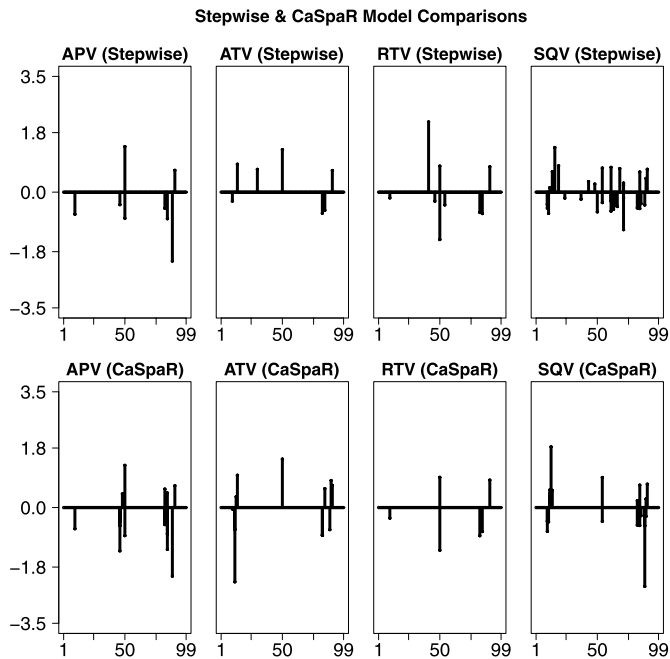}
\vspace*{-3pt}
  \caption{Comparison of stepwise and CaSpaR models
  across four drugs: APV, ATV, RTV and SQV.  Each plot
  gives the coefficients for the selected mutation predictors,
  versus the locations of these mutations in the protein sequence.
  Each
  vertical line represents the magnitude of the coefficient for a
   mutation predictor. Note that some sequence locations can have
  multiple mutations.}\label{4drugfig}
  \vspace*{-6pt}
\end{figure}

We present a summary of our results in Table \ref{HIV}.  Compared to
stepwise regression, CaSpaR has comparable mean-squared-error (MSE) and
number of mutations selected.  In most cases, CaSpaR selects a few more
mutations and has a~slightly lower MSE.  The lasso generally does
better in terms of MSE, but includes many more mutations.  These
results are complicated and cumbersome to interpret as a model of
resistance. Overall, CaSpaR gives relatively sparse models, as desired.

Figure \ref{4drugfig} compares the sparsity structure in the CaSpaR and
stepwise solutions in four of the drugs.  If we compare the sparsity
patterns of the stepwise and CaSpaR solutions, we see that CaSpaR gives
more clustered solutions, as expected.  As mentioned before, CaSpaR and
stepwise regression select about the same number of mutations.  The
clustered CaSpaR solutions, however, select mutations from fewer
positions than stepwise regression.  The CaSpaR models therefore give a~comparable level of prediction accuracy and sparsity, while also having
a~better biological interpretation: these clusters may correspond to a
functional area of the protein.

%s5 ###
\section{Simulation study}

We next report the results of a simulation study.  We show that CaSpaR
recovers a structured sparsity pattern more effectively than forward
stepwise regression and lasso.    For CaSpaR, we use a Stetson kernel,
and tune the parameters with a grid of $h = \{1,2,3,4\}$, and $\alpha =
\{0.1,0.2,\ldots,1\}$.  For each method, we use 10-fold cross-validation
to choose all tuning parameters and stopping times.  To measure the
performance of each method, we use
%
%e9 ###
\begin{equation}
\mbox{Recovery Error} = \frac{\Vert\hat{\beta} -
\overline{\beta}\Vert^2_2}{\Vert\overline{\beta}\Vert^2_2},
\end{equation}
where $\hat{\beta}$ is the coefficient estimated by the method and
$\overline{\beta}$ is the true coefficient vector.  This metric appeals
to us since it captures both selection and estimation performance.  We
also compare the true positive rate and false positive rate in order to
directly measure selection performance.  Here, a true positive is when
a nonzero entry of $\hat{\beta}$ is also nonzero in $\overline{\beta}$.
A false positive is when a nonzero entry of $\hat{\beta}$ is zero in
$\overline{\beta}$.

%f3 ###
\begin{figure}[b]

\includegraphics{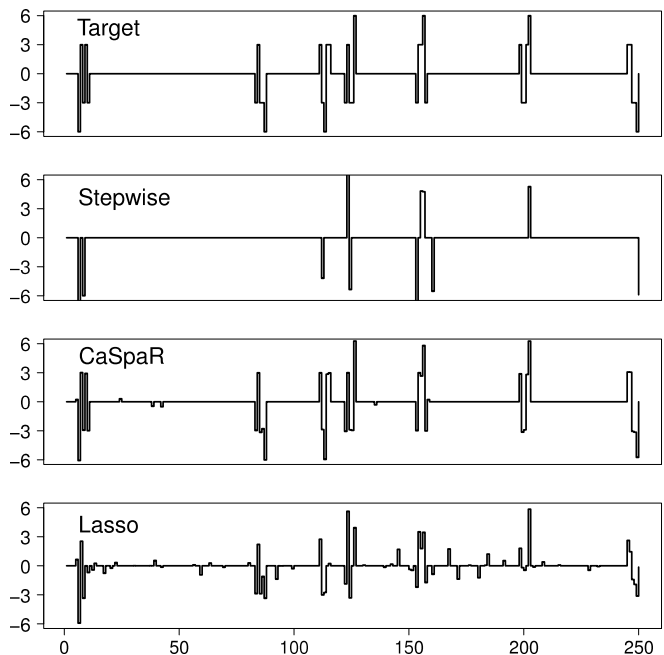}
\vspace*{-3pt}
  \caption{Recovery of coefficients for a single simulated data
  set $(n=100)$.  The top panel displays the target coefficient vector.  The next three panels show the
estimated coefficients for Stepwise, CaSpaR and LASSO, having
recovery errors 0.848, 0.059, 0.542, respectively.}\label{single}
\vspace*{-6pt}
\end{figure}

We simulate 100 $n \times p$ data matrices $\mathbf{X}$ with $p = 250$
columns.  Each entry of these $\mathbf{X}$ is an i.i.d. draw from a
standard normal distribution.  We generate 100 corresponding true
coefficient vectors $\overline{\beta}$ so that each has 7 groups of 5
nonzero coefficients, randomly placed. Thus, there are 35 nonzero
entries in each $\overline{\beta}$.  Within each nonzero group, we set
one entry of $\beta$ equal to 6, and the rest equal to 3 (see the top
panel of Figure \ref{single} for a display of a sample coefficient
vector).  We then randomize the signs of the nonzero entries.  We add
independent Gaussian noise with variance 1 to the simulated response.

\begin{figure}

\includegraphics{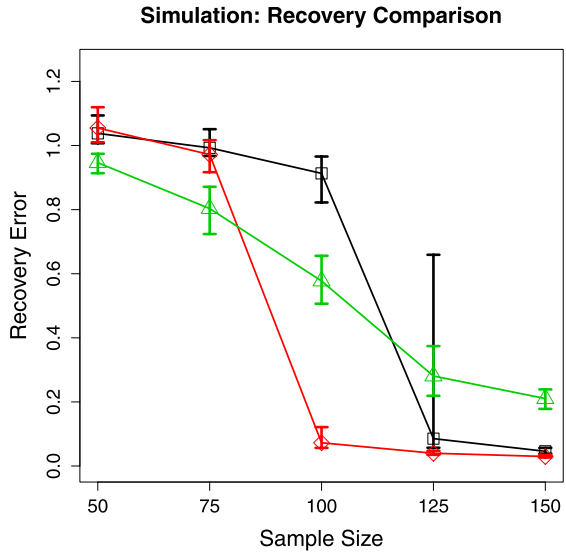}
\vspace*{-3pt}
  \caption{Recovery error $(\Vert\hat{\beta} - \beta\Vert_2^2 /
    \Vert\beta\Vert_2^2 )$ on simulated data with 1-dimensional
  structured sparsity.  Black points: stepwise
  regression; green points: lasso; red points: CaSpaR.  We can see that with less data
CaSpaR achieves a much better recovery rate than either of the other
two methods.}\label{simfig}
\vspace*{-6pt}
\end{figure}

\begin{figure}

\includegraphics{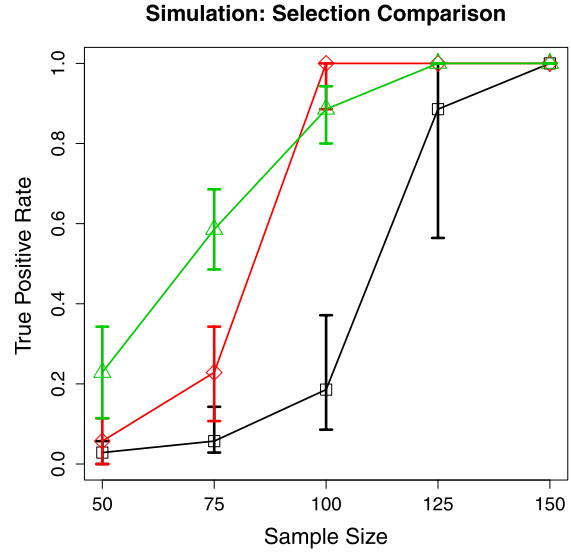}

  \caption{True positive rate (number of correctly identified
  nonzero entries of $\overline{\beta}$ in $\hat{\beta}$/total
  number of nonzero entries of $\overline{\beta}$) on
  simulated data with 1-dimensional
  structured sparsity.  Black points: stepwise
  regression; green points: lasso; red points: CaSpaR.
CaSpaR is competitive with the other two methods.  Note that the
superior true positive rate of the lasso comes at the cost of a high
rate of false positives.} \label{simfig-rec}
\end{figure}

\begin{figure}

\includegraphics{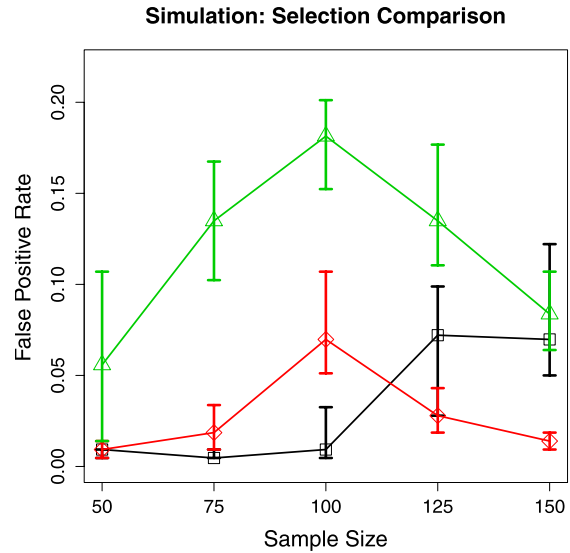}

  \caption{False negative rate (number of incorrect nonzero
  entries of $\hat{\beta}$ with respect to $\overline{\beta}$/total number of nonzero entries of $\overline{\beta}$) on simulated data with 1-dimensional
  structured sparsity.  black points: stepwise
  regression; green points: lasso; red points: CaSpaR.
CaSpaR achieves a superior rate to the lasso.  Note that the y-axis
range is $[0,0.2]$.} \label{simfig-rec-fnr}
\end{figure}

To compare the three methods, we increase $n$ from 50 to 150 ($n =
50,75,\break 100,125,150$) and compare the average recovery errors of
the three methods; cf. Figure~\ref{simfig}. CaSpaR gives near-optimal
performance with fewer data points than the other methods.  An example
of the differences in performance between the three methods on a single
simulated data set ($n=100$) is given in Figure \ref{single}.  CaSpaR
recovers the signal well, while the other two methods do not. Figures
\ref{simfig-rec} and \ref{simfig-rec-fnr} display a comparison of the
true positive rates and false positive rates of the three methods.  We
see that CaSpaR achieves the best balance of these two properties, with
near optimal performance when $n=150$---a property not seen with
stepwise regression or the lasso. We therefore conclude that CaSpaR can
reconstruct sparse signals more effectively than stepwise regression or
the lasso.

%f4 ###

%f5 ###

%f6 ###

%s6 ###
\section{Theoretical properties}

In this section we discuss the theoretical properties of CaSpaR. We
begin by explaining how CaSpaR relates to other methods.

%s6.1 ###
\subsection{Related work}

Several existing regression methods take into account structure as well
as sparsity.  \cite{Yuan06modelselection} introduced the
\textit{grouped lasso}, which allows only groups of predictors to be
selected at once. This is desirable when the groups represent closely
linked predictors---such as a set of predictors that code the levels of
a multilevel factor predictor.  Since this method modifies the lasso,
it can be cast as a convex minimization problem. However, the groups
have to be predefined, and the method does not
 allow for overlap between groups, making this method somewhat
inflexible.

\cite{1553429} introduced an algorithm called StructOMP that modifies
forward stepwise regression (also known as orthogonal matching pursuit
or OMP).  Here, the desired sparsity structure is encoded as a set of
blocks, each of which is assigned a cost.  The algorithm proceeds by
greedily adding blocks one at a time to reduce the loss, scaled by the
cost of the added block. StructOMP allows for very flexible sparsity
structures.  In particular, it can approximate a general class of
sparsity structures the authors term \textit{graph sparsity}, which we
discuss in Section \ref{ss:graphspar}.

Recent work by \cite{1553431} relating to the grouped lasso extends the
possible group structures to include overlapping groups.  Like
StructOMP, the overlapping group penalty can produce models that
approximately follow graph sparsity.  This approach has the advantage
of being a convex minimization problem.  As we discuss in the next
section, for graph sparsity, this method, like StructOMP, gives only an
approximation to graph sparsity because of computational
considerations.

%s6.2 ###
\subsection{Graph sparsity}\label{ss:graphspar}

Graph sparsity is a specific type of structured sparsity introduced by
\cite{1553429}.  Consider a graph $G$ whose nodes include the set
$\mathcal{I} = \{1,2,\ldots,p\}$.  Thus, each predictor is a node of
$G$, but for generality we allow other nodes to be in the graph as
well. We then define the neighborhood of a node $v$ as the set of nodes
with an edge connecting it to $v$.  More generally, we could allow for
$k$-neighborhoods---the set of all nodes with a path of at most $k$
edges connecting it to $v$.  We then consider a sparsity structure
where the important predictors appear within neighborhoods, or a series
of connected neighborhoods.

For example, consider a grid graph, such as in the case of a pixelated
image.  Each pixel is connected to four neighbors, one to each cardinal
direction.  The sparsity structure for this graph connects visually
related components in the image.

CaSpaR can approximate graph sparsity if we employ an appropriate
distance function and bandwidth.  Given a graph $G$, the distance
function can be defined in terms of the graph:
%
%e10 ###
\begin{equation}
d(l,m) = \min\{\mbox{Length of paths from } l \mbox{ to } m,\mbox{
  as defined by }G\}.
\end{equation}

More generally, each edge can be weighted, and $d(\cdot,\cdot)$ can be
the minimal weighted path length.  We then can define neighborhood size
via the bandwidth~$h$.  For the Stetson kernel, the mixing parameter
$\alpha$ controls the number of connected neighborhoods, where $\alpha
= 0$ allows only one. In the image example, we can define
$d(\cdot,\cdot)$ as above.  Then, with $h \in (1,2)$, CaSpaR would find
a sparsity structure of connected pixels.

CaSpaR is a very flexible way to approximate graph sparsity.  First, it
allows for neighborhoods to be locally defined through the bandwidth
while still allowing neighborhoods to grow arbitrarily large as the
method proceeds.  Second, when used with the Stetson kernel, CaSpaR
allows the user to control the degree to which graph sparsity is
adhered via the mixing parameter $\alpha$.

In comparison, the algorithms for the StructOMP of Huang,
Zhang and Metaxas~(\citeyear{1553429}) and \textit{graph lasso} of \cite{1553431} approximate
graph sparsity by constructing a set of node neighborhoods, based on
the graph structure. These generate a set of blocks or groups, that are
then used in the OMP or group lasso framework, respectively. However,
to control the computational cost, they limit the neighborhood size
used to make these blocks or groups. Because CaSpaR grows neighborhoods
instead of seeking to add them all at once as a group or block, this is
not necessary. These algorithms can handle large groups or blocks, but
only at significant computational cost.

Further, in StructOMP, there is no clear way to control the degree to
which graph sparsity is followed in the solution.  The blocks are each
assigned a cost, but this cost is relatively restrictive.  In graph
lasso, the group penalty is controlled by a parameter $\lambda$, just
as with the $\ell_1$ lasso penalty.  However, the group penalty
controls sparsity as well as the structure, so as $\lambda$ decreases,
the model becomes less sparse as well as less structured. A separate
$\ell_1$ penalty could allow the model to be controlled for sparsity
and structure separately.

%s6.3 ###
\subsection{Consistency}

We now explain how a result in \cite{1577088} on stepwise regression
can be adapted to CaSpaR.  We summarize the result from the literature
as follows: under assumptions about the data matrix and the response,
it can be shown that with high probability, when the forward stepwise
procedure stops, it stops with all correctly selected predictors---that
is, all the nonzero entries of the final $\hat{\beta}$ are also
nonzero in the true target $\beta$.  Note that there may be additional
``false negatives.''  Moreover, if all of the target coefficients are
above a threshold set by the noise level, then the entire sparsity
pattern is captured exactly.

We closely follow the proof in \cite{1577088}.  This result requires
more conditions than the similar result for stepwise regression.
However, since we assume that we have a certain set of tuning
parameters $\{\alpha, h\}$, the assumptions are not too harsh.  For
ease of reference, we use notation similar to \cite{1577088}.

We have an $n\times p$ matrix $X$ consisting of $p$ $n$-vectors $\{
\ex_1,\ex_2,\ldots,\ex_p\}$, and an $n$-vector $\mathbf{y}$.   We
assume that there is a target $\overline{\beta} \in \mathbb{R}^p$, such
that
%
%e11 ###
\begin{equation}
\mathbb{E} \mathbf{y} = X\overline{\beta}.
\end{equation}
This assumption means that the linear model is correct. It also roughly
means there is a target coefficient vector $\overline{\beta}$ that
estimates $\mathbf{y}$ well, relative to the noise level.  For both
stepwise and CaSpaR methods, we define $\beta^{(k)}$ as the coefficient
vector after the $k$th step.  Recall the definition of the support of a
vector:
%
%e12 ###
\begin{equation}
\operatorname{supp}(\beta) = \{j\dvtx \beta_j \neq 0 \}.
\end{equation}
We then define $F^{(k)} = \operatorname{supp}(\beta^{(k)})$, $\overline{F} =
\operatorname{supp}(\overline{\beta})$. Let
%
%e13 ###
\begin{equation}
\hat{\beta}_{\mathbf{X}}(F,\mathbf{y}) = \arg \min_{\beta \in
  \mathbb{R}^p} \Vert\mathbf{X}\beta - \mathbf{y}\Vert_2^2\quad
\mbox{subject to}\quad \operatorname{supp}(\beta) \subseteq F.
\end{equation}
Finally, we define two technical quantities:
%
%e14 ###
\begin{equation}
\mu_X(\overline{F}) = \max_{j \notin \overline{F}} \|
(X_{\overline{F}}^TX_{\overline{F}})^{-1}X_{\overline{F}}^T\ex_j\|_1
\end{equation}
and
%
%e15 ###
\begin{equation}
\rho_X(\overline{F}) = \inf_{\beta} \biggl\{ \frac{1}{n}
  \|X\beta\|_2^2/\|\beta\|_2^2 \dvtx \operatorname{supp}(\beta) \subset
  \overline{F} \biggr\}.
\end{equation}
For CaSpaR, we define a distance measure on our predictor index
$1,2,\ldots,p$: $d(\cdot,\cdot)$.  We assume that we are using a~boxcar
kernel, or a Stetson kernel with a~boxcar kernel: $K_{h,m}(l) =
I_{d(md(k,l) < h}.$ We then define the following set, which represents
the candidate predictors---predictors not already included in the model---``underneath'' the
kernel:
%
%e16 ###
\begin{equation}
\mathbb{A}^{(k)} = \bigl\{ m\dvtx d(l,m) < h, m \notin F^{(k)} \bigr\}.
\end{equation}
It follows that
%e17 ###
\begin{equation}
   W_j = \cases{
       \alpha + (1-\alpha)/k \dvtx j \in \mathbb{A}^{(k)},\cr
       \alpha \dvtx j \notin \mathbb{A}^{(k)}.}
\end{equation}

Finally, recall that we have $\epsilon$ as the stopping criterion for
CaSpaR.  If at step $k$ we select $\ex_{i(k)}$ as the next predictor to
be included in the model, then if
%
%e18 ###
\begin{equation}
\bigl|\ex_{i(k)}^T\bigl(X\beta^{(k-1)} - \mathbf{y}\bigr)\bigr| \leq \epsilon,
\end{equation}
CaSpaR stops at step $k-1$.

\begin{theorem}
Suppose that:
\begin{enumerate}
\item $\frac{1}{n}\Vert\ex_j\Vert_2^2 = 1$ $\forall j \in 1,2,\ldots,p$.
\item $\exists \overline{\beta} \in \mathbb{R}^p$, with $\overline{F} =
\operatorname{supp}(\overline{\beta})$ s.t. $\mathbb{\mathbf{y}} =
X\overline{\beta}$.
\item $\mu_X(\overline{F}) < 1$.
\item $\rho_X(\overline{F}) > 0$.
\item The elements of $\mathbf{y}$;
$[y_i]_{i = 1,2,\dots,n}$ are independent sub-Gaussian random
variables: $\exists \sigma>0 \mbox{ s.t. } \forall i, \forall t \in
\mathbb{R}, \mathbb{E} e^{t(y_i - \mathbb{E}y_i)} \leq e^{\sigma^2t^2 /
2}$.
\item Given $\eta \in (0,1)$, let the stopping criterion satisfy
\[
\epsilon > \frac{1}{1 - \mu_X(\overline{F})}
\sigma\sqrt{2\log(2p/\eta)}.
\]
\item There are $\{\alpha,h\}$ such that for each $k$, at least one of
the following conditions holds:
\begin{longlist}
\item[(a)] $\frac{\max_{j \notin \overline{F}} |\ex_{j}^T(X\beta^{(k-1)} -
\mathbf{y})|}{\max_{i \in \overline{F}} |\ex_{i}^T(X\beta^{(k-1)} -
\mathbf{y})|} < \alpha$,
\item[(b)] $\mathbb{A}^{(k-1)} \subseteq
\overline{F}$,
\item[(c)] $\mathbb{A}^{(k-1)} \supseteq \overline{F}$.
\end{longlist}
\end{enumerate}
Then, when the procedure stops at step $k-1$, with probability greater
than $1 - 2\eta$, the following hold:
\begin{enumerate}
\item[1.] $F^{(k-1)} \subset \overline{F}$,
\item[2.] $|\overline{F} -
F^{(k-1)}| \leq 2| \{j \in \overline{F}\dvtx |\overline{\beta}_j| <
3\epsilon \rho_X(\overline{F})^{-1} / \sqrt{n} \}|$,
\item[3.] $\Vert\beta^{(k-1)} - \hat{\beta}_{\mathbf{X}}(\overline{F},\mathbf{y})\Vert_2
\leq \epsilon \rho_X(\overline{F})^{-1}\sqrt{ | \overline{F} -
F^{(k-1)} | / n}$,
\item[4.] $ \Vert\hat{\beta}_{\mathbf{X}}(\overline{F},\mathbf{y}) - \overline{\beta}
\Vert_{\infty} \leq \sigma \sqrt{ 2 \log(2 |\overline{F}/\eta) /
(n\rho_X(\overline{F}))}.$
\end{enumerate}
\end{theorem}

We omit the proof as it is very similar to the proof in \cite{1577088}.

%s6.3.1 ###
\subsubsection{Discussion of the result}

The theorem states that when the procedure stops: (1) the selected
predictors have truly nonzero $\overline{\beta}_i$; (2) the number of
false negatives is bounded by the number of small truly nonzero
$\overline{\beta}_j$---relative to the noise level; (3) the estimator
is close to the best possible $\beta$, which is estimated in the
presence of noise using all the truly nonzero predictors; and (4) the
difference between the best estimate in the presence of noise and that
of the true $\overline{\beta}$ is bounded.

The proof of this result is based on induction at each step of the
procedure.  The extra conditions are motivated by the following
analysis. We denote any predictor for which $\overline{\beta}_j = 0$ as
a \textit{noise} predictor and any predictor for which $\overline{\beta}_j
\neq 0$ as a~\textit{signal} predictor. When we consider adding a
predictor in a step of forward stepwise regression, we consider two
quantities:
%
%e20 ###
%e19 ###
\begin{eqnarray}
&&\max_{j \notin \overline{F}} \bigl|\ex_{j}^T\bigl(X\beta^{(k-1)} - \mathbf{y}\bigr)\bigr|, \\
&&\max_{i \in \overline{F}} \bigl|\ex_{i}^T\bigl(X\beta^{(k-1)} - \mathbf{y}\bigr)\bigr|.
\end{eqnarray}
These are, respectively, proportional to the maximum correlation
between the current residuals and a noise predictor and the maximum
correlation between the current residuals and a signal predictor.  We
refer to these two predictors as the ``best'' signal predictor and the
``best'' noise predictor.

For CaSpaR, we must consider how the weights applied to these
quantities affect the analysis.  We therefore consider the cases where:
(a) the best signal predictor and the best noise predictor are in
$\mathbb{A}^{(k)}$, (b) neither the best signal predictor nor the best
noise predictor are in $\mathbb{A}^{(k)}$, or (c) the best signal
predictor is in $\mathbb{A}^{(k)}$ but the best noise predictor is not,
or (d) the best noise predictor is in $\mathbb{A}^{(k)}$ but the best
signal predictor is not. Except for scenario (d), the original result
for stepwise regression holds.  We therefore make additional
assumptions to ensure that case (d) does not occur.  Those conditions
are as follows:
\begin{enumerate}
\item The ratio of the criterion for the best noise predictor to the
  best signal predictor is less than $\alpha$.
\item All of the predictors under the kernel are signal predictors.
\item All of the signal predictors are under the kernel.
\end{enumerate}
The first ensures that in case (d) the correlation between the signal
predictor is large enough to be selected even in this case.  Because
the weights $W_j$ only depend on membership in $\mathbb{A}^{(k)}$, the
second and third conditions ensure that case (d) never occurs: the
second means there are only signal predictors in~$\mathbb{A}^{(k)}$,
and the third means that there are no signal predictors not in
$\mathbb{A}^{(k)}$.

These assumptions are fairly mild, especially if we have a strong
belief that $\operatorname{supp}(\overline{\beta})$ is truly structured. We
propose that the first condition holds for early steps of CaSpaR. We
can reasonably assume that it is possible for an oracle $\alpha$ to be
such that the signal is sufficiently dominant over noise.  The last two
conditions should hold for later steps of the algorithm: enough points
within each cluster have already been discovered so that it only
remains to fill in the clusters.

%s7 ###
\section{Conclusion}

We introduced a new method, CaSpaR, that allows us to build sparse
regression models where we have some additional information about the
structure of the sparsity pattern.  We presented an application as well
as a simulation study that show the method performs differently than
the most popular sparse regression techniques.  We discussed the
general concept of graph sparsity, and showed that, under high
``signal-to-noise'' conditions $\Vert\beta\Vert^2/\sigma \approx 500$, our
method provides a flexible way to approximate graph sparsity.

Our simulation study suggests that under structured sparsity
conditions, CaSpaR can recover the true target with less data than
standard techniques.  This motivates future work to show that this
property has a theoretical basis.  Other topics of interest include
adding backward steps to the CaSpaR algorithm as well as an extension
to a convex minimization procedure, which may have some computational
advantages over the stepwise procedure.

\section*{Acknowledgments}
The authors would like to thank the reviewers for~ma\-ny helpful
comments.

\printaddresses

\end{document}